\documentclass[preprint,12pt]{elsarticle}

\usepackage{graphicx}
\usepackage{amssymb}

\usepackage{amsmath, amssymb, latexsym}
\usepackage{blindtext, rotating}
\usepackage{tikz}

\journal{Journal Name}

\begin{document}

\begin{frontmatter}


\title{Deep Learning parameterization of subgrid scales in wall-bounded turbulent flows.}

\author{Anikesh Pal}
\ead{pala@ornl.gov}
\address{National Center for Computational Sciences, Oak Ridge National Laboratory, TN USA\fnref{label3}}




\begin{abstract}
An innovative \textit{deep learning} approach has been adopted to formulate the eddy-viscosity for large eddy simulation (LES) of wall-bounded turbulent flows. A deep neural network (DNN) is developed which learns to evaluate the eddy-viscosity from a dataset generated for a channel flow at friction Reynolds number $Re_{\tau} = 395$ using the Dynamic Smagorinsky subgrid-scale model. Later this DNN is employed to predict the eddy-viscosity for a number of grid configurations for channel flow at $Re_{\tau} = 395$ and $590$. The statistics computed from the DNN based LES model show an excellent match with direct numerical simulations (DNS). 
In some cases, particularly for the coarse grid simulations, the DNN based model yields statistics closer to DNS than those from the Dynamic Smagorinsky model. 
The turbulent kinetic energy budget terms also manifest a satisfactory match with the DNS results. This model computes eddy-viscosity $2-8$ times quicker than the Dynamic Smagorinsky model. This DNN based LES model is also able to closely mimic the duct flow at $Re_{\tau} = 300$ in a qualitatively and quantitatively similar manner as the LES using the Dynamic Smagorinsky model and a DNS from a previous study. This study demonstrates the feasibility of deep learning for parameterizing the subgrid-scales (SGS) in a turbulent flow accurately in a cost-effective manner. In a broader perspective, deep learning based models can be a promising alternative to traditional RANS and LES models for simulating complex turbulent flows.\\
\end{abstract}


\end{frontmatter}


\section{Introduction}
Machine learning is a type of artificial intelligence that enables a computer to learn without being explicitly programmed. 
The basic idea for machine learning is to develop a learning algorithm, input the algorithm with training data sets (representing the experience of the algorithm) and experiment the algorithm with unknown testing data sets before it can be deployed to make intelligent decisions for real applications. A subset of machine learning known as deep learning \citep{lecun:2015} has evolved as one of the most compelling and cutting-edge topics of research and has demonstrated tremendous increase in accuracy in the areas of image and speech recognition. Deep learning achieves its great power and flexibility by learning categories incrementally through its hidden layer architecture. Owing to its supremacy in terms of accuracy when trained with a huge amount of data, deep neural networks are gaining popularity in the areas of fluid mechanics and turbulence modeling.\\

The onset of utilizing neural networks (NN) in fluid mechanics trace back to the studies of \cite{lee:1997,giralt:2000,milano:2002,sarghini:2003}. NN based controllers were applied to turbulent channel flow for drag reduction in \cite{lee:1997}. \cite{giralt:2000} presented an artificial NN capable of capturing the basic non-linearity in the turbulent wake of a cylinder. NNs are also used by \cite{milano:2002} to reconstruct the near wall fields in a turbulent channel flow. A NN based subgrid scale LES model was developed by \cite{sarghini:2003}, and was trained on the data generated by LES of a channel flow at $Re_{\tau} = 180$ with Bardina’s scale similar (BFR) SGS model. This model when executed for a channel flow at $Re_{\tau} = 180$, was able to reproduce the highly nonlinear behavior of the flows at a $20\%$ cheaper computational cost than the BFR model.\\

 In the past few years NN based models have emerged as a substitute to turbulent closure schemes in Reynolds Averaged Navier Stokes (RANS) models \citep{tracey:2013,zhang:2015,parish:2016,ling:2016}. The RANS models are widely used in engineering applications owing to their computational effectiveness as compared to DNS and LES. Most of the RANS models ( Spalart-Allmaras \citep{spalart:1992}, $k-\epsilon$ \citep{chien:1982}, SST $k-\omega$ \citep{menter:1994}) are either one or two equation models. The accuracy of these models are dependent on the formulation of the eddy-viscosity to establish a relationship between the Reynolds stresses and the mean strain rate. The computation of the eddy-viscosity for turbulence closure is subjected to certain inherent assumptions and may fail in capturing the Reynolds stress anisotropy for moderately complex flows \citep{tracey:2013}. To improve the prediction of the Reynolds stress anisotropy tensor in RANS models NNs are designed from high-fidelity simulations (DNS, LES) and experimental data \citep{tracey:2013,zhang:2015,parish:2016,ling:2016}. \cite{tracey:2013} developed a regression model and trained it on a dataset from a DNS simulation to predict the anisotropy of the Reynolds stress tensor. When the model was applied to a channel flow, it provided improved prediction of turbulence anisotropy as compared to the $k-\omega$ RANS model. Later, \cite{tracey:2015} used a shallow (single hidden layer) NN to reproduce the source term in the Spalart-Allmaras model for a wide variety of 
flow conditions. Random forest regressors were used in \cite{ling:2017} to predict the Reynolds stress anisotropy in a jet-in-crossflow configuration. These random forests regressors provided significantly improved Reynolds stress anisotropy predictions as compared to the default RANS predictions. But random forest regressors barely preserve the tensor invariance properties \citep{lingb:2016}. Tensor invariance refers to the independence of the properties of a tensor, such as the trace, scalar product and determinant, with respect to coordinate transformation. 
DNNs can identify these invariance properties efficiently owing to their hierarchical learning architecture. \cite{lingb:2016} compared the performance of DNNs and random forest algorithms trained on invariant and raw tensor inputs for predicting the Reynolds stress anisotropy. Both the models performed similarly when trained on invariant tensors. However, for raw tensor inputs the DNNs outshined random forest. \cite{ling:2016} embedded Galilean invariance into a DNN and reported significant improvement in the prediction of the Reynolds stress anisotropy compared to linear and non-linear eddy-viscosity models in RANS models. A review on the recent developments on the use of machine learning to improve turbulence models is provided in \cite {duraisamy:2018}.\\

All these previous studies primarily focused on developing some form of machine learning algorithm to calculate the Reynolds stress anisotropy tensor in RANS models. The present investigation however, uses \textit{deep learning} as an ingenious technique to formulate the eddy viscosity ($\nu_{sgs}$) rather than the SGS stresses in a LES model for channel flow at $Re_{\tau} = 395,590$ and a duct flow at $Re_{\tau} = 300$. \cite{gamahara:2017} developed a shallow NN for modeling the SGS stresses in a turbulent channel flow, however their model was "unable" to show any advantage over the Standard Samgorinsky model. This current LES model is termed as \textit{intelligent} eddy-viscosity (INU) model. Once this model is trained and validated, it replaces the Dynamic Smagorinsky subgrid-scale model for computing the eddy-viscosity in a turbulent channel and duct flow. Comparison of statistics among the INU model, the Dynamic Smagorinsky model and previous DNS results are shown to comment on the fidelity of the INU model. To further appraise the robustness of the INU model, the turbulent kinetic energy budget terms are also assessed for the channel flow.\\

\section{Method}
\label{method}
\subsection{Problem setup}
The details of the computational domain for the channel flow are shown in figure \ref{fig1}(a). The non-dimensional parameter $Re_{\tau}$ resembles the study of Moser, Kim and Mansour \citep{moser:1999} (MKM). Periodic boundary conditions are used in the streamwise $(x)$ and spanwise $(z)$ directions, whereas the top and bottom wall are subjected to no-slip boundary conditions. The simulation is initialized with a uniform flow in the streamwise direction with random fluctuations near the wall.  

\subsection{Governing Equations and numerical method}
In a LES model, the equations of motion for an incompressible flow are filtered in space using a top hat filter, and are written in non-dimensional form as:\\
\begin{equation}
\frac{\partial \overline{u}_i}{\partial x_i} = 0,
\end{equation}
\begin{equation}
\label{governing}
\frac{\partial \overline{u}_i}{\partial t} + \frac{\partial {(\overline{u_j u_i}})}{\partial x_j}= -\frac{\partial \overline{P}}{\partial x_i} + \frac{1}{Re_{\tau}}\frac{\partial^2 \overline{u}_i}{\partial x_j \partial x_j} - \frac{\partial \tau_{ij}}{\partial x_j}, 
\end{equation}
where the overbar denotes the filtered quantities. The subgrid scale stress, $\tau_{ij}$ is parameterized by the Dynamic Smagorinsky model (DSM) \cite{germano:1991} as follows:
\begin{equation}
\tau_{ij} = -2C_d\overline{\Delta}^2|S|S_{ij},    
\end{equation}
where $\overline{\Delta}$ is the filter width, $C_d$ is the model coefficient, $\overline{S}_{ij} = 1/2(\partial \overline{u}_i/\partial x_j + \partial \overline{u}_j/\partial {x_i})$ is the resolved strain rate and $|S|$ is defined as $\sqrt{2\overline{S}_{ij}\overline{S}_{ij}}$. The subgrid eddy-viscosity is given by:\\
\begin{equation}
\label{eqn4}
\nu_{sgs} = C_d\overline{\Delta}^2|S|,    
\end{equation}
where $C_d$ is calculated by a dynamic procedure \citep{germano:1991,lilly:1992} in which a test filter is applied to the resolved velocity fields. The quantities denoted by $\widetilde{\overline{\bullet}}$ are double-filtered with both LES and test filters. The dynamic coefficient is computed by:\\
\begin{equation}
C_d = -\frac{1}{2}\frac{\langle L_{ij}M_{ij}\rangle}{\langle M_{ij}M_{ij}\rangle}, 
\label{cd}
\end{equation}
where $L_{ij} = \widetilde{\overline{u}_i\overline{u}}_j - \Tilde{\overline{u}}_i\Tilde{\overline{u}}_j$ and $M_{ij} = \widetilde{\overline{\Delta}}^2\widetilde{|\overline{S}|}\widetilde{\overline{S}}_{ij} - \widetilde{\overline{\Delta}^2|\overline{S}|\overline{S}}_{ij}$. The ratio of the test and LES filter, $\widetilde{\overline{\Delta}}/\overline{\Delta} = 6$ and $\langle \bullet \rangle$ in equation \ref{cd} denotes averaging in the homogeneous directions. For time-advancement, a semi-implicit, third order Runge-Kutta/Crank-Nicolson formulation is used. The viscous terms in the wall normal direction are treated implicitly whereas the viscous terms in the periodic direction are treated explicitly. All the spatial derivatives are discretized using a central, second-order finite difference scheme on a staggered grid. The pressure Poisson equation which is utilized to project the velocity field into a divergence-free space is solved using a multigrid solver. This DNS and LES solver has been validated and used extensively for a number of free-shear and wall-bounded turbulence problems \citep{pham:2018,pal:2015,pal:2013,brucker:2010}.

\subsection{Deep neural networks}
\label{DNNx}
 A dense, fully-connected, feed-forward DNN is chosen for this application. An example DNN of this type is as follows:
\begin{equation}
\label{eqna}
h_p^1 = \varphi^1\left(\sum_{i=1}^nW_{ip}^1\mathcal{A}_{i} + b_p^1\right);\hspace{0.5cm} p \in [1,p],
\end{equation}
\begin{equation}
\label{eqnb}
h_q^2 = \varphi^2\left(\sum_{j=1}^qW_{jq}^2h_j^1 + b_q^2\right);\hspace{0.5cm} q \in [1,q],
\end{equation}
\begin{equation}
\label{eqnc}
h_r^3 = \varphi^3\left(\sum_{k=1}^rW_{kr}^3h_k^2 + b_r^3\right);\hspace{0.5cm} r \in [1,r],
\end{equation}
\begin{equation}
\label{eqnd}
\mathcal{C}_{out} = \left(\sum_{l=1}^mW_{ml}^4h_m^3 + b_m^4\right);\hspace{0.5cm} m \in [1,m].
\end{equation}
Here $\mathcal{A}(1\times9)$ denotes the input vector; $\mathcal{C}_{out}(1\times1)$ is the predicted output vector; $n = 9$ is the number of inputs, $p, q, r = 32$ are the number of neurons per hidden layer and $m = 1$ is the output, $W^1 (9\times32)$, $W^2(32\times32)$, $W^3(32\times32)$, $W^4(32\times1)$ are matrices of trainable weights; $b^1(1\times32)$, $b^2(1\times32)$, $b^3(1\times32)$, $b^4(1\times1)$ are ``bias vectors''; $\varphi^1$, $\varphi^2$, $\varphi^3$ are the non-linear activation functions; and $h^1(1\times32)$, $h^2(1\times32)$, and $h^3(1\times32)$ are the ``hidden'' vectors whose scalar components are called ``neurons''. Note that activation functions are only applied on the hidden vectors. For the present investigation a deep learning package called Keras\footnote{{h}ttps://keras.io} has been used, which is a high-level wrapper around Tensorflow\footnote{{h}ttps://www.tensorflow.org/} written in python.\\

\subsection{Data collection and normalization}
Table \ref{table1} lists all the simulated cases for this study. DSM1 represents the base case simulated at $Re_{\tau} = 395$ using the Dynamic Smagorinsky model and serves as a source of data collection for a deep neural network (DNN). The domain is decomposed laterally on $64$ CPU cores for computation. The input and output variables are collected at each grid point along the colored dash lines in figure \ref{fig1}(a) from each CPU core at every $100$ time steps. This ensures spatial and temporal variability in the input-output pairs.
The Dynamic Smagorinsky model computes $\nu_{sgs}$ from the velocities, strain rates and the size of the filters (LES and test) at every grid point. Therefore, the velocities $(\overline{u}, \overline{v}, \overline{w})$ and strain rates $(\overline{S}_{ij} = 1/2(\partial \overline{u}_i/\partial x_j + \partial \overline{u}_j/\partial {x_i})$) \citep{germano:1991} at every grid point are taken as inputs to the DNN whereas $\nu_{sgs}$ at the corresponding locations will be the output from the DNN. The size of the filters are related to the grid size and are taken into account during strain rate calculation. Therefore the effect of the size of the filters are indirectly associated with the strain rates and are excluded from the inputs to the DNN. Approximately $77$ million input-output samples are collected from DSM1.\\

\begin{table*}
\centering
\begin{tabular}{lrrrrrrrrrr}
\hline
Case&$N_x$ & $N_y$ & $N_z$ &  $Re_{\tau}$ & $\Delta x^{+}$ & min($\Delta y^{+}$) & max($\Delta y^{+}$) & $\Delta z^{+}$  & CPU time (s)  \\
\hline
DSM1$^*$&128&256&128&395&19.4&0.46&6.4&9.7&0.27\\
INULES1&128&256&128&395&19.4&0.46&6.4&9.7&0.13\\
INULES2&64&128&64&395&38.7&0.91&12.9&19.4&\\
INULES3&48&64&48&395&51.7&1.9&25.7&25.8&0.004\\
DSM3&48&64&48&395&51.7&1.9&25.7&25.8&0.032\\
INULES4&128&256&128&590&28.9&0.68&9.5&14.5&\\
INULES5&64&128&64&590&57.9&1.3&19.2&28.9&0.008\\
DSM5&64&128&64&590&57.9&1.3&19.2&28.9&0.068\\
\hline
\end{tabular}
\caption{Simulation parameters for channel flow: $N_x$, $N_y$, $N_z$ are the number of grid points in x, y and z directions respectively. $Re_{\tau} = \frac{u_{\tau}\delta}{\nu}$ is the friction Reynolds number, where $u_{\tau} = 1m/s$ is the friction velocity, $\delta = 1m$ is the half channel height and $\nu(m^2/s)$ is the molecular viscosity. The grid resolution $\Delta x^{+} = \frac{u_{\tau}\Delta x}{\nu}$, $\Delta y^{+} = \frac{u_{\tau}\Delta y}{\nu}$ and $\Delta z^{+} = \frac{u_{\tau}\Delta z}{\nu}$ are given in terms of wall units $(\frac{\nu}{u_{\tau}})$. The min($\Delta y^{+}$) is the resolution at the wall and max($\Delta y^{+}$) is the resolution at the center of the channel. The CPU time represents the time elapsed for eddy-viscosity calculation in each time step. The data is collected from DSM1$^*$. }
\label{table1}
\end{table*}

The dataset comprising of the input-output pairs are randomly separated into a training ($90\%$ of the collected samples) and testing dataset ($10\%$ of the collected samples). This random splitting assures spatial and temporal variability of the input-output pairs in the training and testing dataset. The training dataset is used to train the DNN whereas the testing data set is used to examine the accuracy of the predictions from the DNN. There are non-uniformities in the values of the input-output variables owing to the fact that they represent variables near the wall and the center of the channel. For example, the streamwise velocity near the wall is a few orders of magnitude smaller than the vertical gradient of the streamwise velocity. These type of incongruities in the input-output pairs can induce huge variation in weight matrices causing numerical instability and impedance to convergence of the optimization algorithm during the training process. Normalizing the data can eradicate the disparity in the input and output variables by scaling them to same order of magnitude resulting in stability and faster convergence of the training process. Each input and output variable in the present study is normalized by the maximum and minimum values of each variable across the entire dataset using the following relations: \\
\begin{equation}
\label{normal1}
\widehat{\mathcal{A}_{ij}} = \frac{\mathcal{A}_{ij} - \min\left(\mathcal{A}_j\right)}{\max\left(\mathcal{A}_j\right) - \min\left(\mathcal{A}_j\right)},
\end{equation} 
\begin{equation}
\label{normal1}
\widehat{\mathcal{C}_{ik}} = \frac{\mathcal{C}_{ik} - \min\left(\mathcal{C}_k\right)}{\max\left(\mathcal{C}_k\right) - \min\left(\mathcal{C}_k\right)},
\end{equation} 
where $\mathcal{A}_{ij}$ are the inputs $(\overline{u},\overline{v},\overline{w},\overline{S}_{xx},\overline{S}_{yy},\overline{S}_{zz},\overline{S}_{xy},\overline{S}_{xz},\overline{S}_{yz})$ and $\mathcal{C}_{ik}$ is the actual output ($\nu_{sgs}$) from DSM1, $i$ is the number of training samples, $j=9$ and $k=1$ . \\

\subsection{DNN setup, training, and validation}
\label{DTIE}
\begin{figure*}
\centering
(a)\includegraphics[height=1.7in]{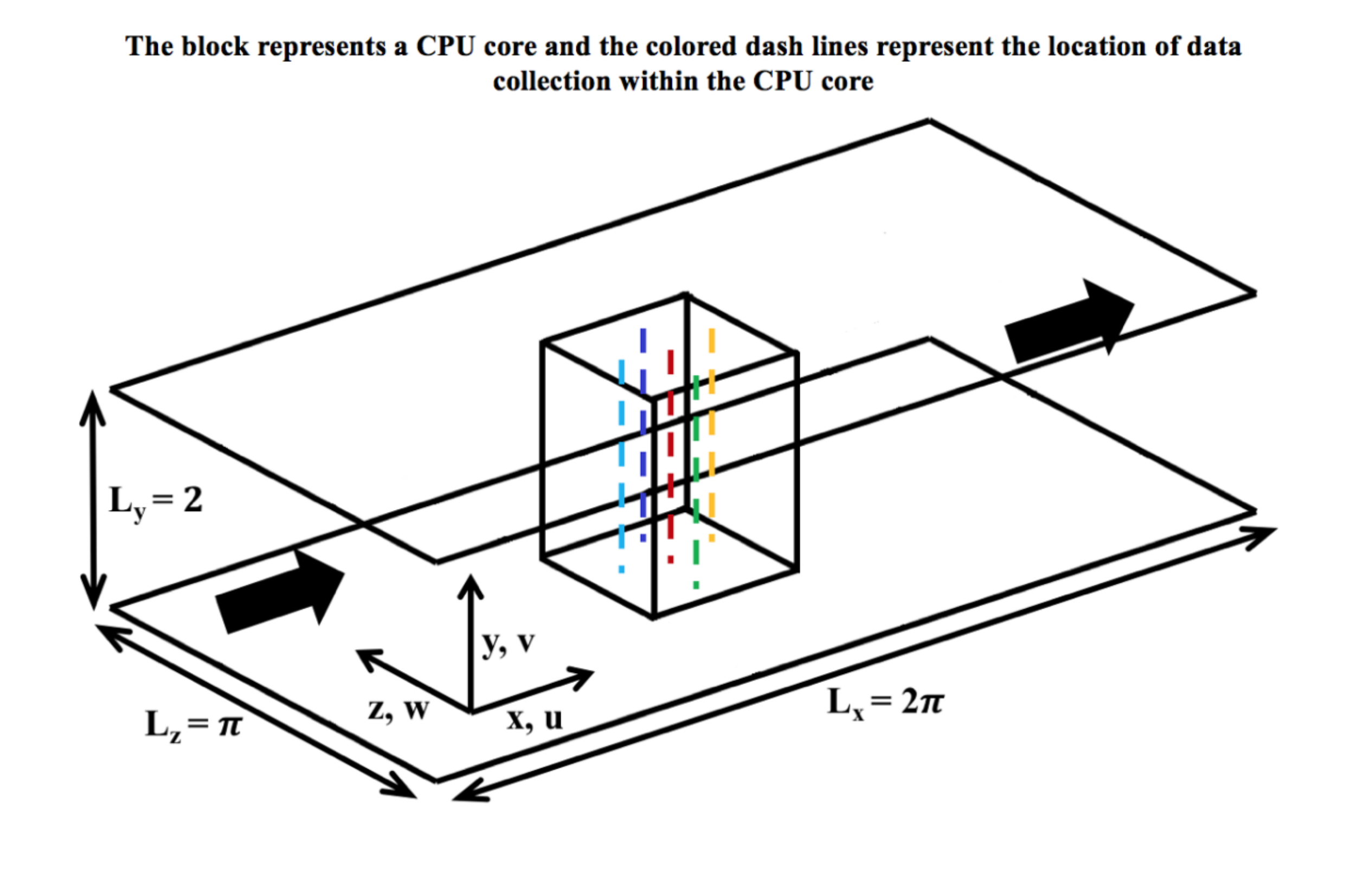}
(b)\includegraphics[height=1.6in]{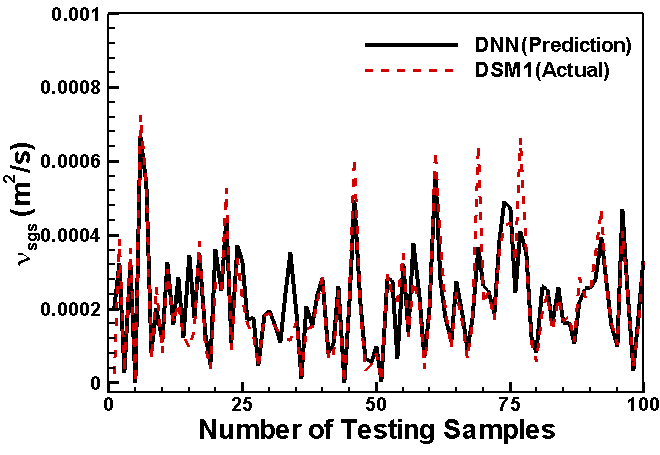}
\caption{(\textit{a}) Schematic for simulating turbulent channel flow. The dimensions ($L_x$, $L_y$, $L_z$) in the respective directions are normalized by the channel half height $\delta = 1m$. (\textit{b}) Comparison of $\nu_{sgs}$ predicted from the DNN with the actual $\nu_{sgs}$ from the Dynamic Smagorinsky Model (DSM1) for the first 100 samples of the testing dataset.}
\label{fig1}
\end{figure*}
Several exploratory training experiments were carried out to refine the number of neurons in each hidden layer. This includes varying the number of hidden layers, number of neurons, different activation functions, different batch sizes and different number of epochs. The goal was to select a number of hidden layers and neurons per layer that will reduce the training loss without affecting the validation error significantly. These goals were achieved by using $3$ "hidden layers" with $32$ neurons in each hidden layer. The normalization performed by equation \ref{normal1} on the input dataset bring the inputs within a range of [0,1]. Owing to the fact that the normalized inputs are positive, a rectified linear unit (ReLU) \footnote{{h}ttps://keras.io/activations/} function defined as $f(\bf{x}) = max(0,\bf{w}\bullet\bf{x})$ is chosen as an activation function for this study. The ReLU activation function provides a slightly better convergence of the training and validation loss as compared to the sigmoid activation function. The optimizer used for training is RMSprop \footnote{{h}ttps://keras.io/optimizers/} which is a variation of stochastic gradient descent (SGD). The DNN is trained with a batch size of $256$ for $100$ epochs \footnote{{h}ttps://keras.io/getting-started/faq/$\#$what-does-sample-batch-epoch-mean} with $mean$ $squared$ $error$ \footnote{{h}ttps://keras.io/losses/} as the loss function. The stopping criterion for training the DNN was determined by the behavior of the training loss and the validation loss. Figure shows that the training loss and the validation loss approach an asymptote after $20$ epochs and do not manifest major variation afterwards. Therefore, $100$ epochs are sufficient for training this DNN. Different combinations of the number of hidden layers, number of neurons and activation functions are explored, however the current configuration is a trade off between the validation accuracy and speed up.

Further details on the setup and architecture of the deep neural network such as hidden layers, number of neurons and activation function are provided in \ref{appendix}. The training took approximately $24$ hours on a single Nvidia Tesla V100-DGXS-16GB graphical processing unit (GPU). Figure \ref{fig1}(b) compares the $\nu_{sgs}$ predicted by the DNN with the actual values of $\nu_{sgs}$ obtained from the Dynamic Smagorinsky model (DSM1) for the first 100 samples of the testing dataset. These samples represent the spatial and temporal variability of $\nu_{sgs}$. The DNN is able to capture the qualitative and quantitative details of the variation in $\nu_{sgs}$ from the Dynamic Smagorinsky model. This DNN will now be referred to as the \textit{intelligent} eddy-viscosity (INU) model and will replace the Dynamic Smagorinsky model for channel and duct flow simulations. The rest of the samples in the testing dataset also manifest qualitative and quantitative similarity in $\nu_{sgs}$ calculated from the DNN and DSM1, however, to avoid data clutter in a single plot only the first 100 samples are shown in figure \ref{fig1}(b).\\
 
\section{Results and discussion}
All the variables are appropriately non-dimensionalized by the friction velocity $u_{\tau}$($m/s$), the channel half height $\delta$($m$) and the molecular viscosity $(\nu)$($m^2/s$).\\
\label{results}
\begin{figure}[ht!]
\centering
(a)\includegraphics[height=1.1in]{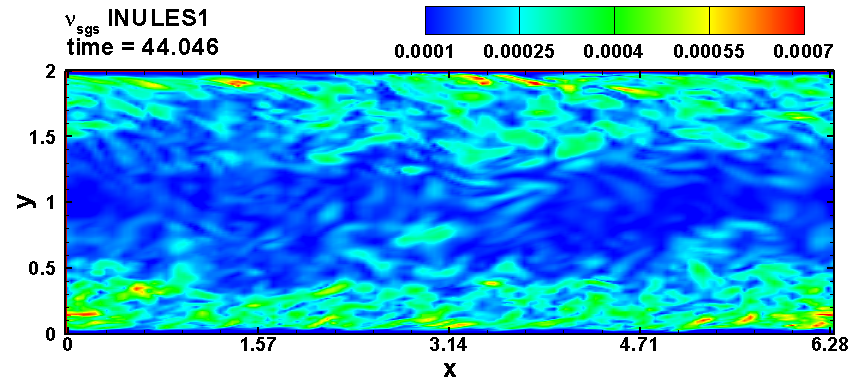}
(b)\includegraphics[height=1.1in]{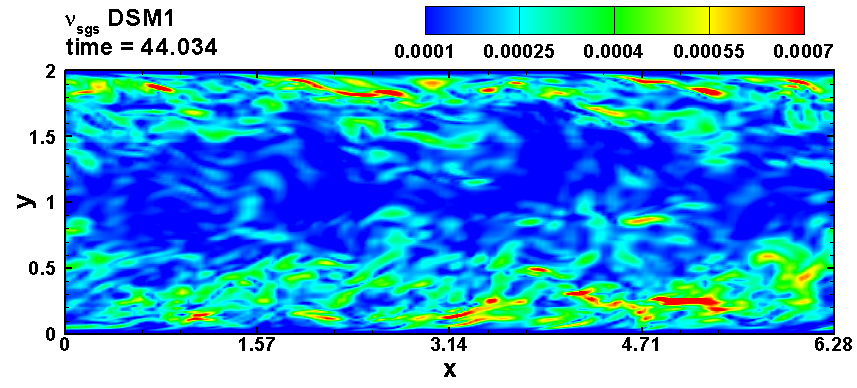}\\
(c)\includegraphics[height=1.1in]{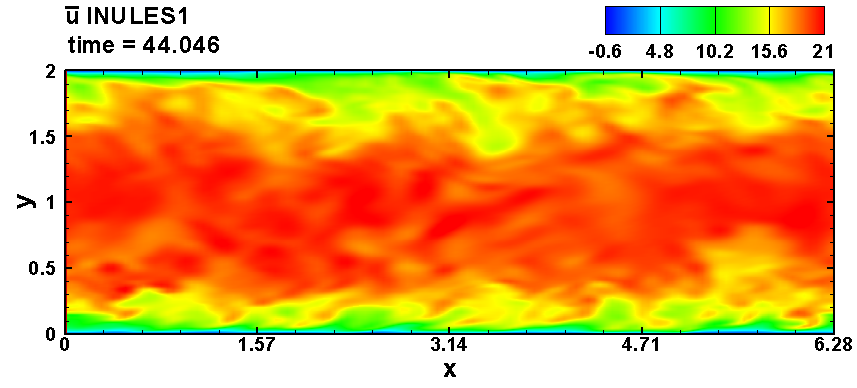}
(d)\includegraphics[height=1.1in]{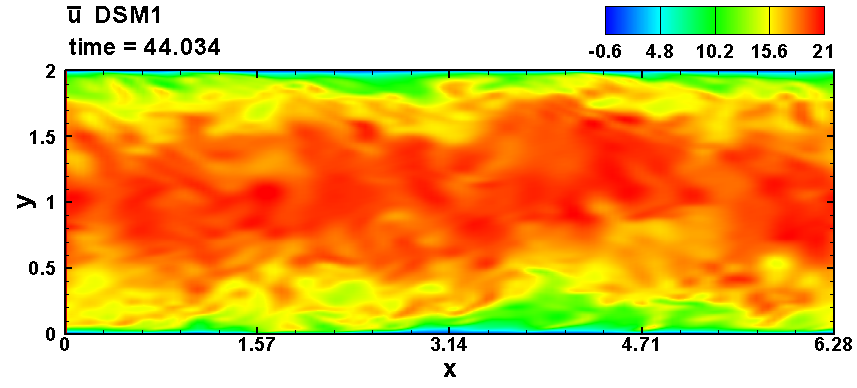}\\
\caption{Contours of instantaneous (\textit{a}) $\nu_{sgs}/\nu$ using INU model, (\textit{b}) $\nu_{sgs}/\nu$ using Dynamic Smagorinsky model, (\textit{c}) $\overline{u}/u_{\tau}$ using INU model, and (\textit{d}) $\overline{u}/u_{\tau}$ using Dynamic Smagorinsky model at a $z = \pi/2$ (x-y) plane for channel flow at $Re_{\tau} = 395$. The time $t$ is normalized by $\delta/u_{\tau}$.}
\label{fig1b}
\end{figure}

Figures \ref{fig1b}(a) and (b) compare the contours of normalized instantaneous eddy-viscosity $(\nu_{sgs}/\nu)$ obtained from the INU model (INULES1) and the Dynamic Smagorinsky model (DSM1) in a vertical plane located at $z = \pi/2$. An analogous comparison of the instantaneous resolved normalized streamwise velocity $(\overline{u}/u_{\tau})$ between the two models are shown in figure \ref{fig1b}(c) and (d). The INU model is able to mimic the distribution of normalized $\nu_{sgs}$ and $\overline{u}$ in a similar manner as the Dynamic Smagorinsky model. \\

\begin{figure}[ht!]
\centering
(a)\includegraphics[height=1.2in]{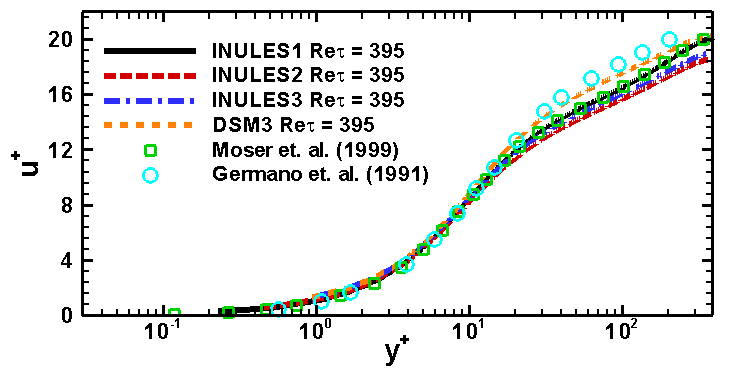}
(d)\includegraphics[height=1.2in]{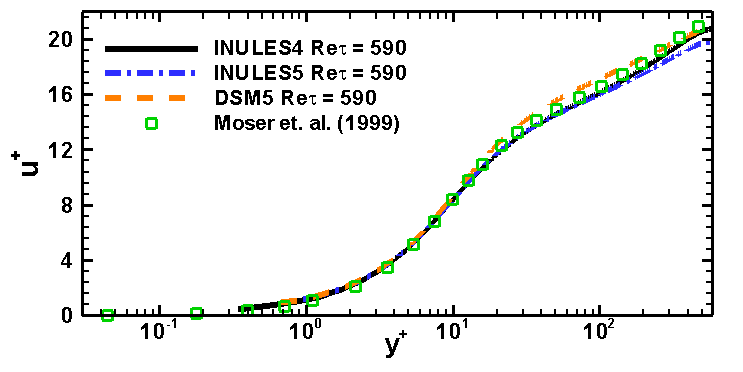}
(b)\includegraphics[height=1.2in]{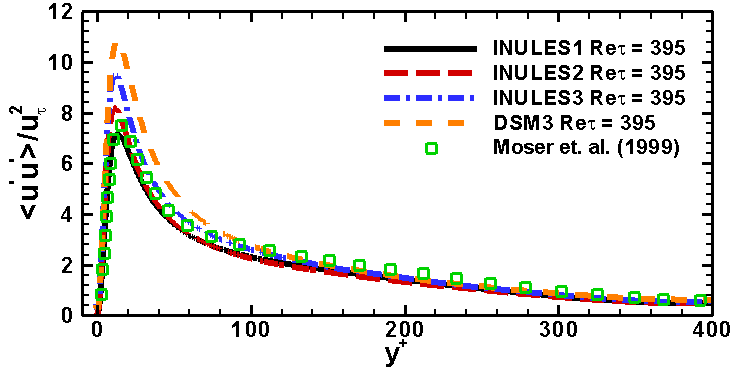}
(e)\includegraphics[height=1.2in]{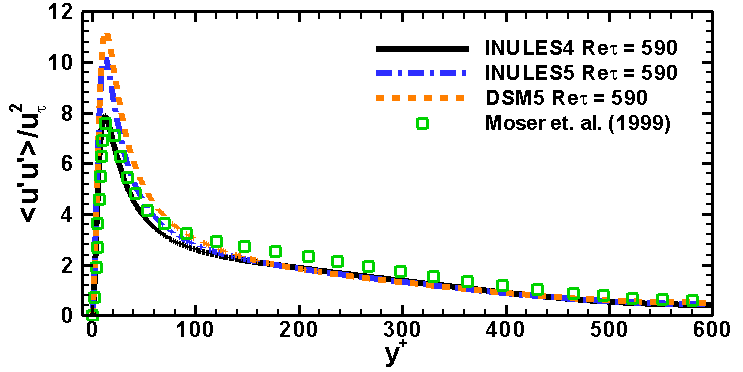}
(c)\includegraphics[height=1.2in]{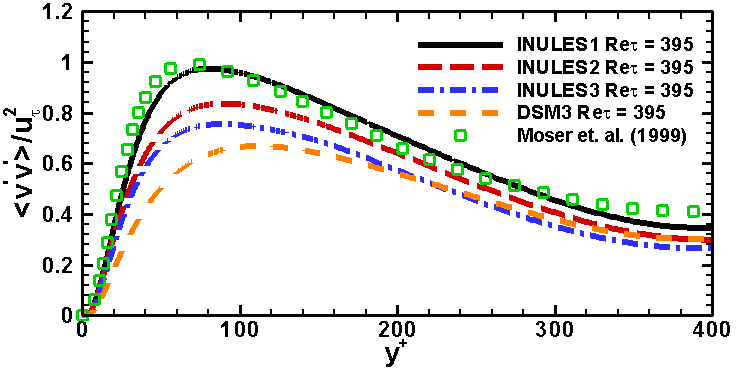}
(f)\includegraphics[height=1.2in]{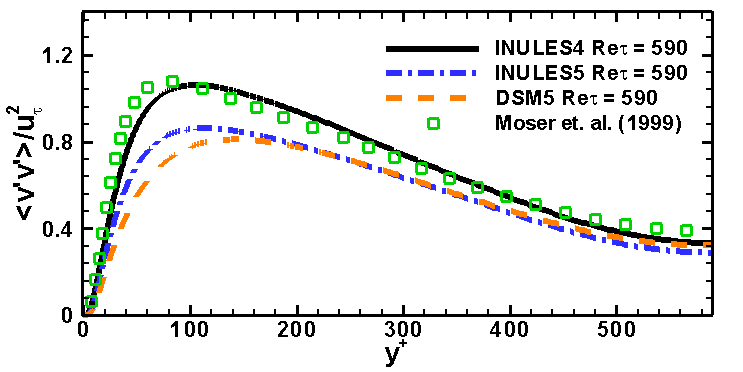}
\caption{Comparison of (\textit{a}) $u^{+}$, (\textit{b}) $\langle u^{'}u^{'} \rangle/u_{\tau}^2$, and (\textit{c}) $\langle v^{'}v^{'} \rangle/u_{\tau}^2$ for channel flow at $Re_{\tau} = 395$. Comparison of (\textit{d}) $u^{+}$, (\textit{e}) $\langle u^{'}u^{'} \rangle/u_{\tau}^2$, (\textit{f}) $\langle v^{'}v^{'} \rangle/u_{\tau}^2$ for channel flow at $Re_{\tau} = 590$.}
\label{fig2}
\end{figure}
A comparison of the statistics with DNS data of \cite{moser:1999} at $Re_{\tau} = 395$ is performed to further examine the fidelity of the INU model. 
The profiles of resolved streamwise mean velocity $\langle \overline{u} \rangle$ normalized by the friction velocity $u_{\tau}$ $(u^{+} = \frac{\langle \overline{u} \rangle}{u_{\tau}})$ is shown as a function of vertical wall coordinates $y^{+}$ in figure \ref{fig2}(a). The simulations with the INU model (INULES1, INULES2 and INULES3) show very good agreement with DNS (\cite{moser:1999}, \footnote{{h}ttp://turbulence.ices.utexas.edu/data/MKM/chan395/profiles/})
and LES (\cite{germano:1991}) in the viscous and the buffer layers. In the outer layer $(y^{+} > 50)$, the $u^{+}$ from the INU model at a relatively fine grid (INULES1) resemble DNS values, whereas the coarse (INULES2, INULES3) grid simulations perform fairly well. Notice that $u^{+}$ for the coarsest (INULES3) grid closely imitate DNS values in the outer layer as compared to slight deviations in $u^{+}$ with respect to the DNS data in the LES of \cite{germano:1991}. The LES of \cite{germano:1991} was performed at a very similar resolution as INULES3. The Reynolds stresses normalized by $u_{\tau}^2$ as a function of $y^{+}$ are shown in \ref{fig2}(b) and (c). The maximum value of $\langle u^{\prime}u^{\prime}\rangle$/$u_{\tau}^2$ obtained from the INU model for the coarsest grid is closer to the DNS results as compared to the Dynamic Smagorinsky model at an analogous resolution (DSM3). Similarly, $\langle v^{\prime}v^{\prime}\rangle$/$u_{\tau}^2$ from the INU model show closer proximity to DNS values as compared to the Dynamic Smagorinsky model for the coarsest grid (DSM3). \\

The INU model also accelerated the computation of $\nu_{sgs}$ by $2-8$ times as compared to the Dynamic Smagorinsky model. This acceleration of the INU model is attributed to the fact that instead of computing $L_{ij}$, $M_{ij}$ and the spatial averages associated with equation \ref{cd}, the DNN calculates $\nu_{sgs}$ by using an optimized matrix multiplier $dgemm$\footnote{http://www.netlib.org/lapack/}. This speed-up is measured by comparing the minimum and maximum CPU time elapsed in calculating $\nu_{sgs}$ among all the processors during the entire simulation for the INU model and the Dynamic Smagorinsky model for the cases (INULES1 $\&$ DSM1; INULES3 $\&$ DSM3; INULES5 $\&$ DSM5) listed in table \ref{table1}.\\

Albeit the INU model is developed from a dataset at $Re_{\tau} = 395$, its generality is tested by simulating channel flow  at $Re_{\tau} = 590$ at fine (INULES4) and coarse (INULES5) grid resolutions. Another LES (DSM5) at a similar grid resolution as INULES5 is also performed by using the Dynamic Smagorinsky model. A comparison of $u^{+}$ and Reynolds stresses normalized by $u_{\tau}^2$ among INULES4, INULES5 and DSM5 with DNS data at $Re_{\tau} = 590$ (\cite{moser:1999},\footnote{{h}ttp://turbulence.ices.utexas.edu/data/MKM/chan590/profiles/})
is presented in figure \ref{fig2}(d)-(f). 
The INU model performs very satisfactorily in the viscous and buffer layer similar to DNS, as evident from the profiles of $u^{+}$ shown in figure \ref{fig2}(d). The DNN based LES model also captures the log-law in the core of the channel very effectively even for the coarse grid simulation. The profiles of $\langle u^{'}u^{'} \rangle/u_{\tau}^2$ and $\langle v^{'}v^{'} \rangle/u_{\tau}^2$ from the INU model also manifest similarity with DNS (figure \ref{fig2}(e)-(f)). Once again the INU model performs better than the Dynamic Smagorinksy model for the coarse grids in terms of the maximum values of $\langle u^{'}u^{'} \rangle/u_{\tau}^2$ and $\langle v^{'}v^{'} \rangle/u_{\tau}^2$. Notice that the DNS \citep{moser:1999} at $Re_{\tau} = 590$ was carried out a resolution of $384\times257\times384$ with $\Delta x^+ = 9.7$, $min(\Delta y^+) = 0.04$, $max(\Delta y^+) = 7.2$ and $\Delta z^+ = 4.8$. Despite the fact that the present simulations are performed at resolutions $9$(INULES4)$-72$(INULES5) times coarser than DNS, the INU model captures all the statistics successfully. \\

\begin{figure}
\centering
(a)\includegraphics[height=1.2in]{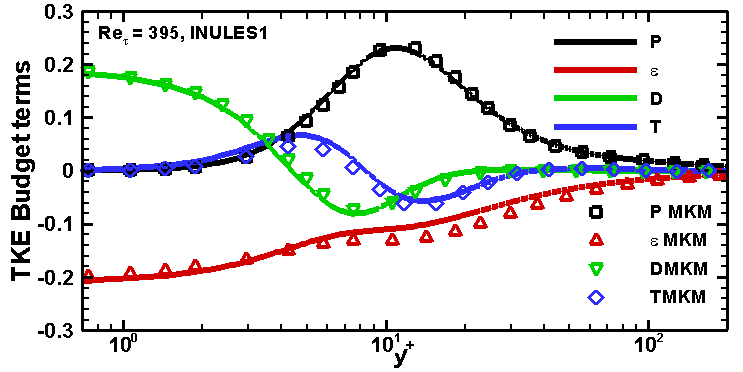}
(d)\includegraphics[height=1.2in]{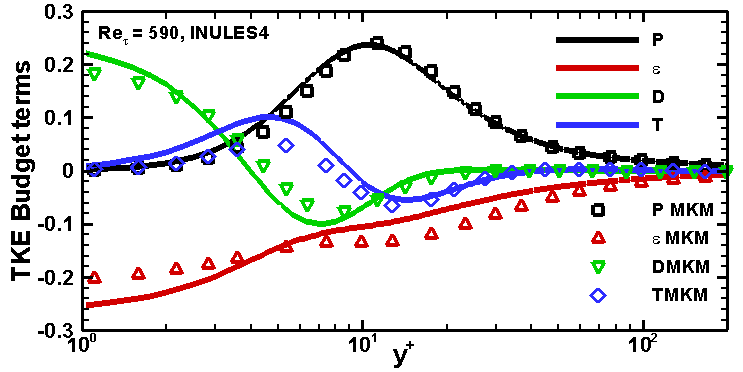}
(b)\includegraphics[height=1.2in]{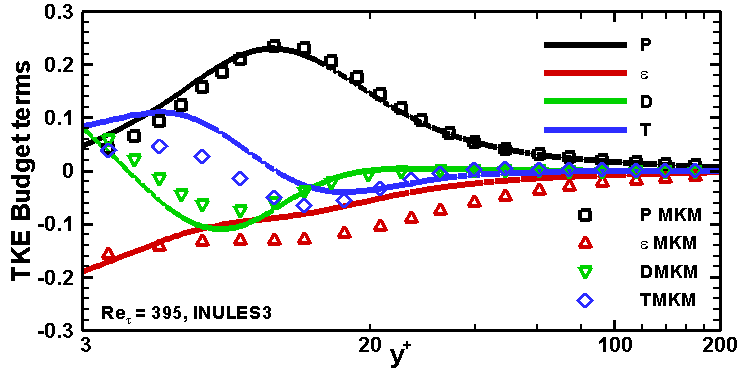}
(e)\includegraphics[height=1.2in]{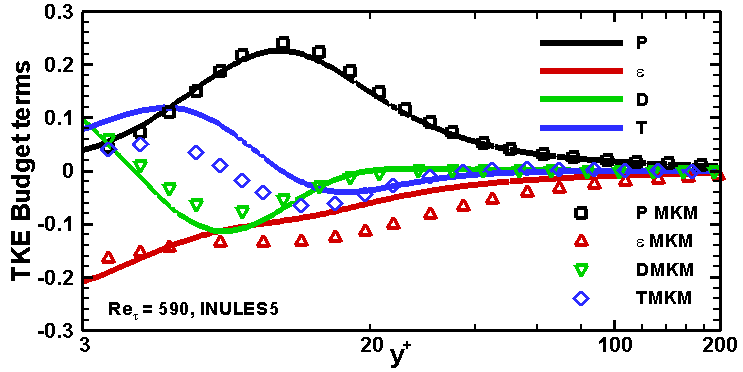}
(c)\includegraphics[height=1.2in]{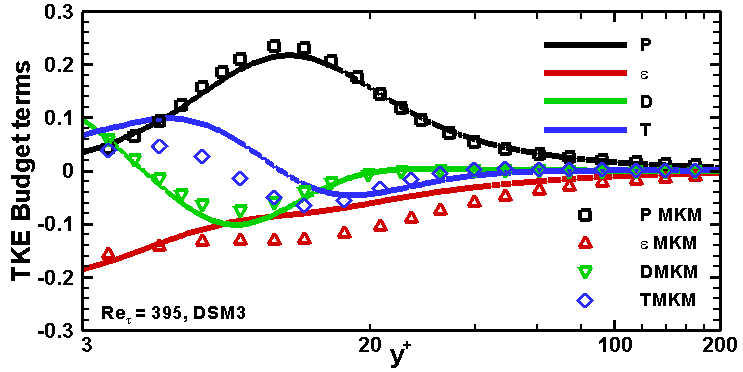}
(f)\includegraphics[height=1.2in]{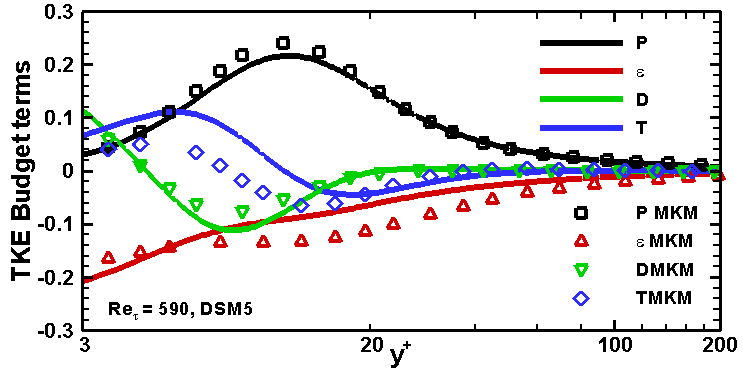}
\caption{Comparison of turbulent kinetic energy budget for channel flow at (\textit{a}) $Re_{\tau} = 395$, INULES1, (\textit{b}) $Re_{\tau} = 395$, INULES3, (\textit{c}) $Re_{\tau} = 395$, DSM3, (\textit{d}) $Re_{\tau} = 590$, INULES4, (\textit{e}) $Re_{\tau} = 590$, INULES5, and (\textit{f}) $Re_{\tau} = 590$, DSM5. MKM is Ref. \cite{moser:1999}.}
\label{fig3}
\end{figure}
A final and more stringent verification of the accuracy of the INU model is the comparison of the turbulent kinetic energy budget terms with the DNS data of \cite{moser:1999}. 
For a fully developed channel flow, the turbulent kinetic energy $(t.k.e)$ equation including the sub-grid scale terms is normalized by $\frac{u_{\tau}^4}{\nu}$ \citep{schiavo:2017} and is written as \citep{pope_2000}:
\begin{align}
0 = \mathcal{P} - \varepsilon - \mathcal{T} - \Pi + \mathcal{D} - \mathcal{T}_{sgs}, 
\end{align}
where 
$\mathcal{P} = -\langle \overline{u}^{\prime}\overline{v}^{\prime}\rangle\frac{d\overline{u}}{dy}/\frac{u_{\tau}^4}{\nu}$ is the production, $\varepsilon$ is the sum of the pseudo-dissipation $(\nu\langle\frac{\partial \overline{u}_i^{\prime}}{\partial y}\frac{\partial \overline{u}_i^{\prime}}{\partial y}\rangle)/\frac{u_{\tau}^4}{\nu}$ and the subgrid scale dissipation $(-\langle \tau_{iy}\frac{\partial \overline{u}_i^{\prime}}{\partial y}\rangle)/\frac{u_{\tau}^4}{\nu}$, $\mathcal{T} = \frac{d\langle\frac{1}{2} \overline{v}^{\prime}\overline{u}_{i}^{\prime}\overline{u}_{i}^{\prime}\rangle}{dy}/\frac{u_{\tau}^4}{\nu}$ is the turbulent transport of $t.k.e$, $\Pi = \frac{1}{\rho}\frac{d\langle \overline{v}^{\prime}\overline{p}^{\prime}\rangle}{dy}/\frac{u_{\tau}^4}{\nu}$ is the pressure transport, $\mathcal{D} = \nu\frac{d^2\langle \frac{1}{2}\overline{u}_i^{\prime}\overline{u}_i^{\prime}\rangle}{dy^2}/\frac{u_{\tau}^4}{\nu}$ is the viscous transport of $t.k.e$ and $\mathcal{T}_{sgs} = \frac{d\langle \tau_{iy}^{\prime}\overline{u}_i^{\prime}\rangle}{dy}/\frac{u_{\tau}^4}{\nu}$ is the subgrid transport. 
The pressure transport $\Pi$, and the subgrid transport $\mathcal{T}_{sgs}$, are very small as compared to the other dominating terms and are not shown to avoid confusion. \\

The profiles of production and viscous transport of $t.k.e$ as a function of $y^+$ from the INU model agree very well with their corresponding DNS profiles for all the cases (figure \ref{fig3}(a)-(f)). The dissipation and turbulent transport profiles for relatively fine grid (INULES1, INULES4) simulations at $Re_{\tau} = 395$ and $590$ agree with their counterpart DNS profiles as shown in figures \ref{fig3}(a) and (d). For the coarse grid (INULES3, INULES5) simulations as shown in figures \ref{fig3}(b) and (e), the dissipation and turbulent transport profiles manifest a descent match with the DNS profiles. The coarse grid LES with the Dynamic Smargorinsky model (DSM3, DSM5) (figures \ref{fig3}(c) and (f)) also show similar dissipation and turbulent transport as the INU model. Therefore the differences between the INU model and DNS are attributed to the coarse grid resolution, and is not a shortcoming of the model itself. \\

LES of a duct flow (computational domain same as \cite{gavrilakis:1992}) is also carried out at $Re_{\tau} = 300$ using the INU model. This simulation further assess the performance of the INU model for a geometry other than a channel. The results of this LES are compared with the DNS of \cite{gavrilakis:1992} at same $Re_{\tau}$.  
The number of grid points for LES is $64 \times 64 \times 64$, which corresponds to $\Delta x^{+} = 147.2$ and $0.7 < \Delta y^{+}, \Delta z^{+} < 9.7$. Note that the number of grids for LES is $64$ times less than the DNS ($1000 \times 127 \times 127$ corresponding to $\Delta x^{+} = 9.4$ and $0.45 < \Delta y^{+}, \Delta z^{+} < 4.6$) of \cite{gavrilakis:1992}.\\

\begin{figure*}
\centering 
(a)\includegraphics[height=1.8in]{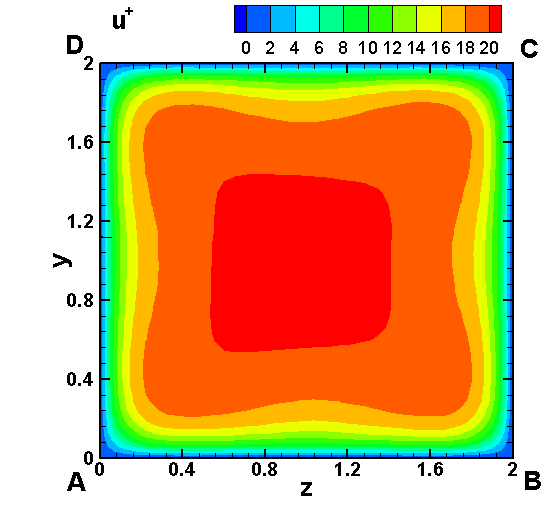}
(b)\includegraphics[height=1.8in]{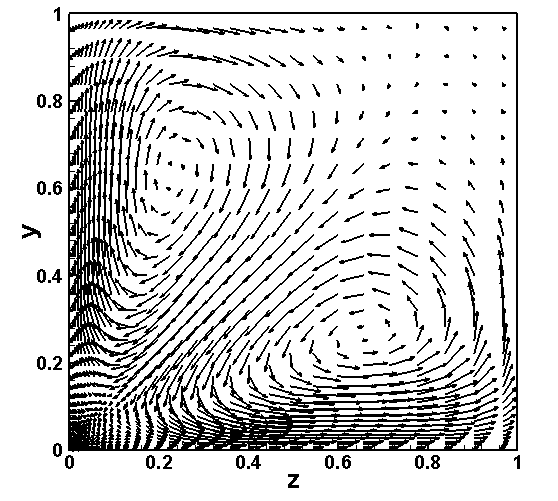}\\
(c)\includegraphics[height=1.2in]{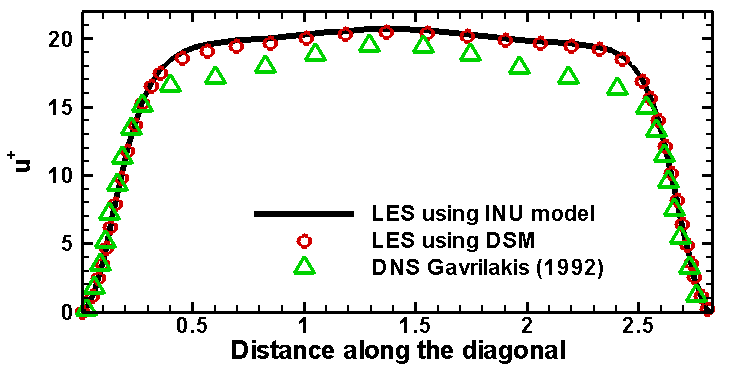}
(d)\includegraphics[height=1.2in]{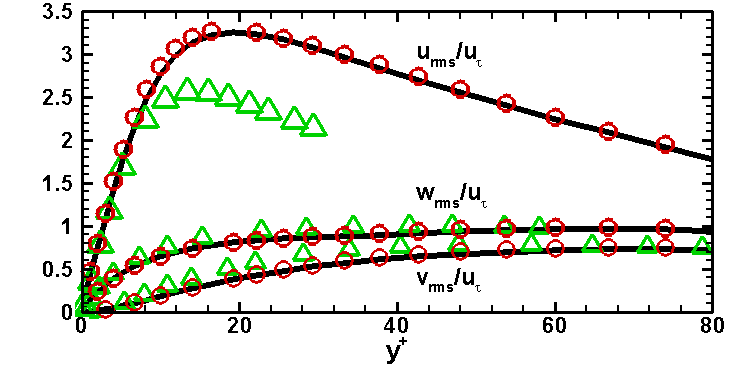}\\
\caption{(\textit{a}) Contours of $u^{+}$, (\textit{b}) mean secondary velocity vectors at the left bottom quadrant, comparison of the profiles of (\textit{c}) $u^{+}$ along the diagonal AC, and (\textit{d}) $r.m.s.$ velocities normalized by $u_{\tau}$ vs $y^{+}$ for a duct flow using the INU model at $Re_{\tau} = 300$.}
\label{fig4}
\end{figure*}
The LES using the INU model successfully captures the distortions of the iso-contours of mean streamwise velocity $u^{+}$ (figure \ref{fig4}(a)) near the corners of the duct similar to \cite{gavrilakis:1992}. These distortions in the mean streamwise velocity are attributed to the momentum transfer by the secondary velocities. These secondary motions are well resolved by the INU model and are shown in figure \ref{fig4}(b) as a vector plot. The INU model calculates $\nu_{sgs}$ for this duct flow simulation $4-5$ times quicker than the Dynamic Smagorinsky model. The $u^{+}$ (\ref{fig4}(c)), and the normalized $r.m.s$ velocities (\ref{fig4}(d)) obtained from the INU model are in perfect accord with the results from the Dynamic Smagorinsky model. The near wall statistics from the INU model match very well with the DNS data \cite{gavrilakis:1992}. The differences in $u_{rms}/u_{\tau}$ between the INU model (as well as the Dynamic Smagorinsky model) and the DNS are an upshot of coarse grid resolution of the LES as compared to the DNS.\\
\section{Conclusions} 
\label{summary}
In this study, a \textit{deep learning} algorithm is developed to formulate an intelligent eddy-viscosity $(INU)$ model for simulating wall bounded turbulent flows. 
The INU model computes the eddy-viscosity from the velocities and the strain rates via a deep neural network. Although the INU model was developed from a dataset obtained from LES of a channel flow at $Re_{\tau} = 395$, it yields results comparable to a DNS for a channel flow at a higher $Re_{\tau} = 590$ . It has been demonstrated that the statistics computed from the INU model resembles closely to DNS data for grids relatively coarser than the DNS itself. The INU model also performs $2-8$ faster than the Dynamic Smagorinsky model in calculating the eddy-viscosity. A separate LES using this INU model is also carried out for a duct flow to appraise its performance for a geometry other than a channel flow. The INU model produced flow features and statistics very similar to the Dynamic Smagorinsky model, and a DNS reported in the literature.\\

This investigation is an initial step towards the inception of $deep$ $learning$ in large eddy simulations. This study has demonstrated that deep neural networks are capable of parameterizing the sub-grid scales accurately in a cost effective manner, and therefore has tremendous potential in the modeling of turbulent flows. There are a lot of opportunities for the development of a universal INU model, but challenges also abound. An immediate follow-up task is to extend this idea to create an INU model from a dataset consisting of samples from fundamental turbulent flows such as jets, shear layers, wakes and wall-bounded flows at high Reynolds numbers. This exercise will expand the applicability of the INU model for general turbulent flows and will build a platform for developing more sophisticated deep learning models for simulating complex flows using LES.\\


\section{acknowledgments}
This research used resources of the Oak Ridge Leadership Computing Facility, which is a DOE Office of Science User Facility supported under Contract DE-AC$05$-$00$OR$22725$. The code for training the DNN is available at \textbf{https$://$github.com$/$itsanikesh$/$DNNLESMODEL}. This repository also contains the weights matrices and bias vectors.\\








\section{Appendix}
\label{appendix}
\subsection*{Method}
\begin{figure}[!ht]
\centering
	\centering
	\begin{tikzpicture}[thick]
		\tikzstyle{unit}=[draw,shape=circle,fill = cyan,minimum size=1.0cm]
		\tikzstyle{hidden}=[draw,shape=circle,fill = cyan,minimum size=1.0cm]

		\node[unit](x1) at (-9.0,9){$\mathcal{A}_1$};
		\node[unit](x2) at (-9.0,7.5){$\mathcal{A}_2$};
		\node[unit](x3) at (-9.0,6){$\mathcal{A}_3$};
		\node at (-9.0,4.5){\vdots};
		\node[unit](xd2) at (-9.0,3){$\mathcal{A}_{n-2}$};
		\node[unit](xd1) at (-9.0,1.5){$\mathcal{A}_{n-1}$};
		\node[unit](xd) at (-9.0,0){$\mathcal{A}_n$};

		\node[hidden](h10) at (-5.0,10.5){$h_1^{(1)}$};
		\node[hidden](h11) at (-5.0,9.0){$h_2^{(1)}$};
		\node[hidden](h12) at (-5.0,7.5){$h_3^{(1)}$};
		\node[hidden](h13) at (-5.0,6){$h_4^{(1)}$};
		\node at (-5.0,4.5){\vdots};
		\node[hidden](h1m3) at (-5.0,3.0){$h_{p-3}^{(1)}$};
		\node[hidden](h1m2) at (-5.0,1.5){$h_{p-2}^{(1)}$};
		\node[hidden](h1m1) at (-5.0,0){$h_{p-1}^{(1)}$};
		\node[hidden](h1m) at (-5.0,-1.5){$h_{p}^{(1)}$};
		
		\node[hidden](h20) at (-1.0,10.5){$h_1^{(2)}$};
		\node[hidden](h21) at (-1.0,9.0){$h_2^{(2)}$};
		\node[hidden](h22) at (-1.0,7.5){$h_3^{(2)}$};
		\node[hidden](h23) at (-1.0,6){$h_3^{(2)}$};
		\node at (-1.0,4.5){\vdots};
		\node[hidden](h2m3) at (-1.0,3.0){$h_{q-2}^{(2)}$};
		\node[hidden](h2m2) at (-1.0,1.5){$h_{q-2}^{(2)}$};
		\node[hidden](h2m1) at (-1.0,0){$h_{q-1}^{(2)}$};
		\node[hidden](h2m) at (-1.0,-1.5){$h_{q}^{(2)}$};

		\node[hidden](h30) at (3.0,10.5){$h_1^{(3)}$};
		\node[hidden](h31) at (3.0,9.0){$h_2^{(3)}$};
		\node[hidden](h32) at (3.0,7.5){$h_3^{(3)}$};
		\node[hidden](h33) at (3.0,6){$h_3^{(3)}$};
		\node at (3.0,4.5){\vdots};
		\node[hidden](h3m3) at (3.0,3.0){$h_{r-2}^{(3)}$};
		\node[hidden](h3m2) at (3.0,1.5){$h_{r-2}^{(3)}$};
		\node[hidden](h3m1) at (3.0,0){$h_{r-1}^{(3)}$};
		\node[hidden](h3m) at (3.0,-1.5){$h_{r}^{(3)}$};

		\node[unit](y1) at (6.0,4.5){$C_{out}$};

        \draw[->] (x1) -- (h10);
		\draw[->] (x1) -- (h11);
		\draw[->] (x1) -- (h12);
		\draw[->] (x1) -- (h13);
		\draw[->] (x1) -- (h1m3);
		\draw[->] (x1) -- (h1m2);
		\draw[->] (x1) -- (h1m1);
		\draw[->] (x1) -- (h1m);

        \draw[->] (x2) -- (h10);
		\draw[->] (x2) -- (h11);
		\draw[->] (x2) -- (h12);
		\draw[->] (x2) -- (h13);
		\draw[->] (x2) -- (h1m3);
		\draw[->] (x2) -- (h1m2);
		\draw[->] (x2) -- (h1m1);
		\draw[->] (x2) -- (h1m);

        \draw[->] (x3) -- (h10);
		\draw[->] (x3) -- (h11);
		\draw[->] (x3) -- (h12);
		\draw[->] (x3) -- (h13);
		\draw[->] (x3) -- (h1m3);
		\draw[->] (x3) -- (h1m2);
		\draw[->] (x3) -- (h1m1);
		\draw[->] (x3) -- (h1m);

        \draw[->] (xd2) -- (h10);
		\draw[->] (xd2) -- (h11);
		\draw[->] (xd2) -- (h12);
		\draw[->] (xd2) -- (h13);
		\draw[->] (xd2) -- (h1m3);
		\draw[->] (xd2) -- (h1m2);
		\draw[->] (xd2) -- (h1m1);
		\draw[->] (xd2) -- (h1m);
		
		\draw[->] (xd1) -- (h10);
		\draw[->] (xd1) -- (h11);
		\draw[->] (xd1) -- (h12);
		\draw[->] (xd1) -- (h13);
		\draw[->] (xd1) -- (h1m3);
		\draw[->] (xd1) -- (h1m2);
		\draw[->] (xd1) -- (h1m1);
		\draw[->] (xd1) -- (h1m);
		
		\draw[->] (xd) -- (h10);
		\draw[->] (xd) -- (h11);
		\draw[->] (xd) -- (h12);
		\draw[->] (xd) -- (h13);
		\draw[->] (xd) -- (h1m3);
		\draw[->] (xd) -- (h1m2);
		\draw[->] (xd) -- (h1m1);
		\draw[->] (xd) -- (h1m);

		\draw[->] (h10) -- (h20);
		\draw[->] (h10) -- (h21);
		\draw[->] (h10) -- (h22);
		\draw[->] (h10) -- (h23);
		\draw[->] (h10) -- (h2m3);
		\draw[->] (h10) -- (h2m2);
		\draw[->] (h10) -- (h2m1);
		\draw[->] (h10) -- (h2m);

        \draw[->] (h11) -- (h20);
		\draw[->] (h11) -- (h21);
		\draw[->] (h11) -- (h22);
		\draw[->] (h11) -- (h23);
		\draw[->] (h11) -- (h2m3);
		\draw[->] (h11) -- (h2m2);
		\draw[->] (h11) -- (h2m1);
		\draw[->] (h11) -- (h2m);

        \draw[->] (h12) -- (h20);
		\draw[->] (h12) -- (h21);
		\draw[->] (h12) -- (h22);
		\draw[->] (h12) -- (h23);
		\draw[->] (h12) -- (h2m3);
		\draw[->] (h12) -- (h2m2);
		\draw[->] (h12) -- (h2m1);
		\draw[->] (h12) -- (h2m);
		
		\draw[->] (h13) -- (h20);
		\draw[->] (h13) -- (h21);
		\draw[->] (h13) -- (h22);
		\draw[->] (h13) -- (h23);
		\draw[->] (h13) -- (h2m3);
		\draw[->] (h13) -- (h2m2);
		\draw[->] (h13) -- (h2m1);
		\draw[->] (h13) -- (h2m);
		
		\draw[->] (h1m3) -- (h20);
		\draw[->] (h1m3) -- (h21);
		\draw[->] (h1m3) -- (h22);
		\draw[->] (h1m3) -- (h23);
		\draw[->] (h1m3) -- (h2m3);
		\draw[->] (h1m3) -- (h2m2);
		\draw[->] (h1m3) -- (h2m1);
		\draw[->] (h1m3) -- (h2m);

        \draw[->] (h1m2) -- (h20);
		\draw[->] (h1m2) -- (h21);
		\draw[->] (h1m2) -- (h22);
		\draw[->] (h1m2) -- (h23);
		\draw[->] (h1m2) -- (h2m3);
		\draw[->] (h1m2) -- (h2m2);
		\draw[->] (h1m2) -- (h2m1);
		\draw[->] (h1m2) -- (h2m);
		
		\draw[->] (h1m1) -- (h20);
		\draw[->] (h1m1) -- (h21);
		\draw[->] (h1m1) -- (h22);
		\draw[->] (h1m1) -- (h23);
		\draw[->] (h1m1) -- (h2m3);
		\draw[->] (h1m1) -- (h2m2);
		\draw[->] (h1m1) -- (h2m1);
		\draw[->] (h1m1) -- (h2m);
		
		\draw[->] (h1m) -- (h20);
		\draw[->] (h1m) -- (h21);
		\draw[->] (h1m) -- (h22);
		\draw[->] (h1m) -- (h23);
		\draw[->] (h1m) -- (h2m3);
		\draw[->] (h1m) -- (h2m2);
		\draw[->] (h1m) -- (h2m1);
		\draw[->] (h1m) -- (h2m);
		
		\draw[->] (h20) -- (h30);
		\draw[->] (h20) -- (h31);
		\draw[->] (h20) -- (h32);
		\draw[->] (h20) -- (h33);
		\draw[->] (h20) -- (h3m3);
		\draw[->] (h20) -- (h3m2);
		\draw[->] (h20) -- (h3m1);
		\draw[->] (h20) -- (h3m);

        \draw[->] (h21) -- (h30);
		\draw[->] (h21) -- (h31);
		\draw[->] (h21) -- (h32);
		\draw[->] (h21) -- (h33);
		\draw[->] (h21) -- (h3m3);
		\draw[->] (h21) -- (h3m2);
		\draw[->] (h21) -- (h3m1);
		\draw[->] (h21) -- (h3m);

        \draw[->] (h22) -- (h30);
		\draw[->] (h22) -- (h31);
		\draw[->] (h22) -- (h32);
		\draw[->] (h22) -- (h33);
		\draw[->] (h22) -- (h3m3);
		\draw[->] (h22) -- (h3m2);
		\draw[->] (h22) -- (h3m1);
		\draw[->] (h22) -- (h3m);
		
		\draw[->] (h23) -- (h30);
		\draw[->] (h23) -- (h31);
		\draw[->] (h23) -- (h32);
		\draw[->] (h23) -- (h33);
		\draw[->] (h23) -- (h3m3);
		\draw[->] (h23) -- (h3m2);
		\draw[->] (h23) -- (h3m1);
		\draw[->] (h23) -- (h3m);

        \draw[->] (h2m3) -- (h30);
		\draw[->] (h2m3) -- (h31);
		\draw[->] (h2m3) -- (h32);
		\draw[->] (h2m3) -- (h33);
		\draw[->] (h2m3) -- (h3m3);
		\draw[->] (h2m3) -- (h3m2);
		\draw[->] (h2m3) -- (h3m1);
		\draw[->] (h2m3) -- (h3m);
		
        \draw[->] (h2m2) -- (h30);
		\draw[->] (h2m2) -- (h31);
		\draw[->] (h2m2) -- (h32);
		\draw[->] (h2m2) -- (h33);
		\draw[->] (h2m2) -- (h3m3);
		\draw[->] (h2m2) -- (h3m2);
		\draw[->] (h2m2) -- (h3m1);
		\draw[->] (h2m2) -- (h3m);
		
		\draw[->] (h2m1) -- (h30);
		\draw[->] (h2m1) -- (h31);
		\draw[->] (h2m1) -- (h32);
		\draw[->] (h2m1) -- (h33);
		\draw[->] (h2m1) -- (h3m3);
		\draw[->] (h2m1) -- (h3m2);
		\draw[->] (h2m1) -- (h3m1);
		\draw[->] (h2m1) -- (h3m);
		
		\draw[->] (h2m) -- (h30);
		\draw[->] (h2m) -- (h31);
		\draw[->] (h2m) -- (h32);
		\draw[->] (h2m) -- (h33);
		\draw[->] (h2m) -- (h3m3);
		\draw[->] (h2m) -- (h3m2);
		\draw[->] (h2m) -- (h3m1);
		\draw[->] (h2m) -- (h3m);

		\draw[->] (h30) -- (y1);

		\draw[->] (h31) -- (y1);

		\draw[->] (h32) -- (y1);
        \draw[->] (h33) -- (y1);
        
        \draw[->] (h3m3) -- (y1);
		\draw[->] (h3m2) -- (y1);

		\draw[->] (h3m1) -- (y1);

		\draw[->] (h3m) -- (y1);
		
		\draw [->          ] (6.6,4.5) -- (7.5,4.5);
		
		\node[text width=3.0cm, anchor=east, right] at (-9.5,11.2) {\bf{Input layer\\ ($n=9$)}};
		\node[text width=4cm, anchor=east, right] at (-8.25,10.25) {\bf${w_{ip}^1, b_p^1, \varphi^1}$};
		\node[text width=3.5cm, anchor=east, right] at (-6.8,12) {\bf{$1^{st}$ hidden layer\\ ($p=32$ neurons)}};
		\node[text width=4cm, anchor=east, right] at (-4.0,11) {\bf${w_{jq}^2, b_q^2, \varphi^2}$};
		\node[text width=3.5cm, anchor=east, right] at (-2.8,12) {\bf{$2^{nd}$ hidden layer\\ ($q=32$ neurons)}};
		\node[text width=4cm, anchor=east, right] at (-0.0,11) {\bf${w_{kr}^3, b_r^3, \varphi^3}$};
		\node[text width=3.5cm, anchor=east, right] at (1.2,12) {\bf{$3^{rd}$ hidden layer\\ ($r=32$ neurons)}};
		\node[text width=4cm, sloped, anchor=east, right] at (5.15,7.0) {\bf${w_{ml}^4, b_l^4}$};
		\node[text width=3.2cm, anchor=east, right] at (4.5,8.0) {\bf{Output layer, 1}};

	\end{tikzpicture}
	\caption{Deep neural network architecture for the INU model.}
    \label{fig6}
\end{figure}
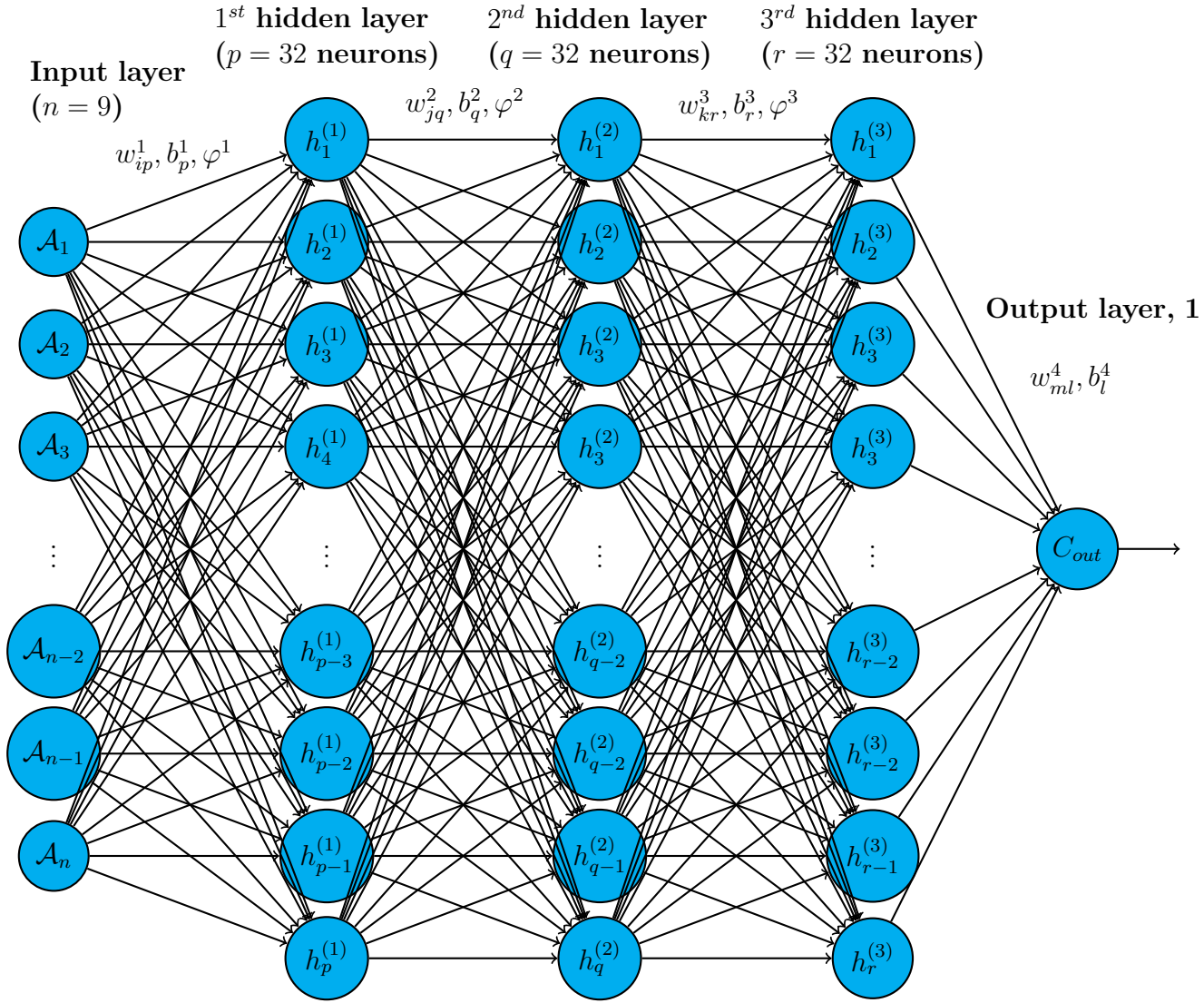

Figure \ref{fig6} shows the deep neural network (DNN) used for this study. It consists of $3$ fully connected layers with $32$ neurons in each layer. The velocities $(\overline{u},\overline{v},\overline{w})$ and the strain rates $(\overline{S}_{xx},\overline{S}_{yy},\overline{S}_{zz},\overline{S}_{xy},\overline{S}_{xz},\overline{S}_{yz})$ constitute the input vector $\mathcal{A}$, $h^1,h^2,h^3$ are the neurons in the $1^{st}, 2^{nd}$ \& $3^{rd}$ hidden layers and $C_{out} = \nu_{sgs}$ is the output. The weights, biases and activation functions for the corresponding layers are represented as $w, b$ \& $\varphi$ respectively. The dimensions of the weight matrix $w^1$ is $9\times32$, $w^2$ is $32\times32$, $w^3$ is $32\times32$ and $w^4$ is $32\times1$. Similarly the dimensions of bias vector $b^1$ is $1\times32$, $b^2$ is $1\times32$, $b^3$ is $1\times32$ and $b^4$ is $1\times1$. The activation function used in this study in a ReLU function defined as $f(\bf{x}) = max(0,\bf{w}\bullet\bf{x})$. The optimizer used for training is RMSprop which is a variation of stochastic gradient descent (SGD). 

\end{document}